\documentclass[prb,aps,showpacs,tightenlines,twocolumn]{revtex4}
\usepackage{graphicx,bm,epsfig}
\usepackage{amssymb}
\usepackage{psfrag}

\begin{document}

\title{Creep of current-driven domain-wall lines: intrinsic
versus extrinsic pinning}

\author{R.A. Duine}
\email{duine@phys.uu.nl} \homepage{http://www.phys.uu.nl/~duine}

\affiliation{Institute for Theoretical Physics, Utrecht
University, Leuvenlaan 4, 3584 CE Utrecht, The Netherlands}

\author{C. Morais Smith}
\email{c.demorais@phys.uu.nl}
\homepage{http://www.phys.uu.nl/~demorais}

\affiliation{Institute for Theoretical Physics, Utrecht
University, Leuvenlaan 4, 3584 CE Utrecht, The Netherlands}

\date{\today}

\begin{abstract}
We present a model for current-driven motion of a magnetic
domain-wall line, in which the dynamics of the domain wall is
equivalent to that of an overdamped vortex line in an anisotropic
pinning potential. This potential has both extrinsic contributions
due to, e.g., sample inhomogeneities, and an intrinsic
contribution due to magnetic anisotropy. We obtain results for the
domain-wall velocity as a function of current for various regimes
of pinning. In particular, we find that the exponent
characterizing the creep regime depends strongly on the presence
of a dissipative spin transfer torque. We discuss our results in
the light of recent experiments on current-driven domain-wall
creep in ferromagnetic semiconductors, and suggest further
experiments to corroborate our model.
\end{abstract}


\maketitle

\def\bx{{\bf x}}
\def\bk{{\bf k}}
\def\bK{{\bf K}}
\def\bq{{\bf q}}
\def\br{{\bf r}}
\def\half{\frac{1}{2}}
\def\args{(\bx,t)}

\section{Introduction} \label{sec:introduction} The driven motion of line defects
through a disordered potential landscape has attracted
considerable attention, for example in the context of vortices in
superconductors,\cite{blatter1994} wetting phenomena,
\cite{degennes1985}, crack fronts, \cite{bouchaud1993} and domain
walls in ferromagnets. \cite{lemerle1998,krusin2001} The
competition and interplay among the elasticity of the line, the
pinning forces due to the disorder potential, and thermal
fluctuations, lead to a wealth of physical phenomena. Topics
discussed are, for example, the universality class of the
roughening of the line, the nature of the pinning-depinning
transition at zero temperature, \cite{kardar1998} and the slide,
depinning, and creep regimes of motion of the line that occur for
decreasing driving field. \cite{blatter1994,feigelman1989}

The creep regime has been observed experimentally with
field-driven motion of domain walls in ferromagnets.
\cite{lemerle1998,krusin2001} This low-field regime is
characterized by a nonlinear dependence of the domain-wall drift
velocity $\langle \dot X \rangle$ on the external magnetic field
$H_{\rm ext}$, given by
\begin{equation}
\label{eq:creepformula} \langle \dot X \rangle \propto \exp
\left\{ - \frac{E_c}{k_B T} \left(\frac{H_c}{H_{\rm ext}}
\right)^{\mu_f} \right\},
\end{equation} where $E_c$ is a characteristic energy scale, and $H_c$ a
critical field. The thermal energy is denoted by $k_B T$ and the
exponent $\mu_f=(2 \zeta -1)/(2-\zeta)$ is given in terms of the
equilibrium wandering exponent $\zeta$ of the static line.
\cite{blatter1994,feigelman1989,lemerle1998} The phenomenological
creep formula [Eq.~(\ref{eq:creepformula})] is an Arrhenius law in
which the energy barrier diverges for vanishing driving field,
i.e., is ``glassy". The underlying assumption is that there is a
characteristic length scale that determines the displacement of
the domain wall line. The validity of Eq.~(\ref{eq:creepformula})
has been confirmed both numerically \cite{kolton2005} and with
functional renormalization group methods. \cite{chauve2000} It
turns out that Eq.~(\ref{eq:creepformula}) is also valid in
situations where roughening plays no role. For example, for a
$d$-dimensional manifold driven through a periodic potential in
$d+1$ dimensions we have $\mu_f = d-1$ (for $d \geq 2$).
\cite{blatter1994} Moreover, in the regime where the line defect
moves via variable-range hopping, we have that $\mu_f=1/3$ if the
motion is in two dimensions.
\cite{blatter1994,nelson1993,thompson1997}

In addition to magnetic-field driven motion, a lot of recent
theoretical and experimental research has been devoted to
manipulating domain walls with electric current
\cite{tatara2004,zhang2004,waintal2004,grollier2003,tsoi2003,yamaguchi2004,
klaui2005,beach2006,hayashi2007,yamanouchi2006} via so-called spin
transfer torques.
\cite{slonczewski1996,berger1996,tsoi1998,myers1999} Domain wall
motion driven by a current is quite distinct from the field-driven
case. For example, it has been theoretically predicted that, in
certain regimes of parameters, the domain wall is intrinsically
pinned at zero temperature, meaning that there exists a nonzero
critical current even in the absence of disorder.
\cite{tatara2004} In clean samples, the phenomenology of
current-driven domain wall motion turns out to crucially depend on
the ratio of the dissipative spin transfer torque
parameter\cite{footnote1} $\beta$  and the Gilbert damping
constant $\alpha_G$. \cite{zhang2004} Although theoretical
predictions \cite{tserkovnyak2006,kohno2006,duine2007,piechon2006}
indicate that, at least for model systems, this ratio differs from
one, it turns out to be difficult to extract its precise value
from experiments on current-driven domain wall motion, to a large
extent because disorder and nonzero-temperature effects
\cite{duine2006,yamanouchi2006} complicate theoretical
calculations of the domain wall drift velocity for a given
current. This is the first motivation for the work presented in
this paper.

Previous work on current-driven domain wall motion at nonzero
temperature focused on {\it rigid} domain walls. Tatara {\it et
al.} \cite{tatara2005} found that $\ln \langle \dot X \rangle$ was
proportional to the current density $j$. Discrepancy of this
result with experiments, \cite{yamanouchi2006} that did not
observe linear dependence of wall velocity on current, motivated
the more systematic inclusion of nonzero-temperature effects on
rigid-domain wall motion by Duine {\it et al.}, \cite{duine2006}
who found that $ \ln  \langle \dot X \rangle \propto \sqrt{j}$ in
certain regimes. Although the latter was an important step in
qualitatively understanding the experimental results of Yamanouchi
{\it et al.}, \cite{yamanouchi2006,yamanouchi2007} a detailed
understanding of these experiments is still lacking and this is
the second motivation of this paper. For completeness, we mention
also the theoretical work by Martinez {\it et al.}
\cite{martinez2007a,martinez2007b} who considered
thermally-assisted current-driven rigid domain wall motion in the
regime of large anisotropy, where the chirality of the domain wall
plays no role and the pinning is essentially dominated by
extrinsic effects. Furthermore, Ravelosona {\it et al.}
\cite{ravelosona2005} have observed thermally-assisted domain wall
depinning,  and Laufenberg {\it et al.} \cite{laufenberg2006} have
determined the temperature dependence of the critical current for
depinning the domain wall.

\begin{figure}
\vspace{-0.5cm} \centerline{\epsfig{figure=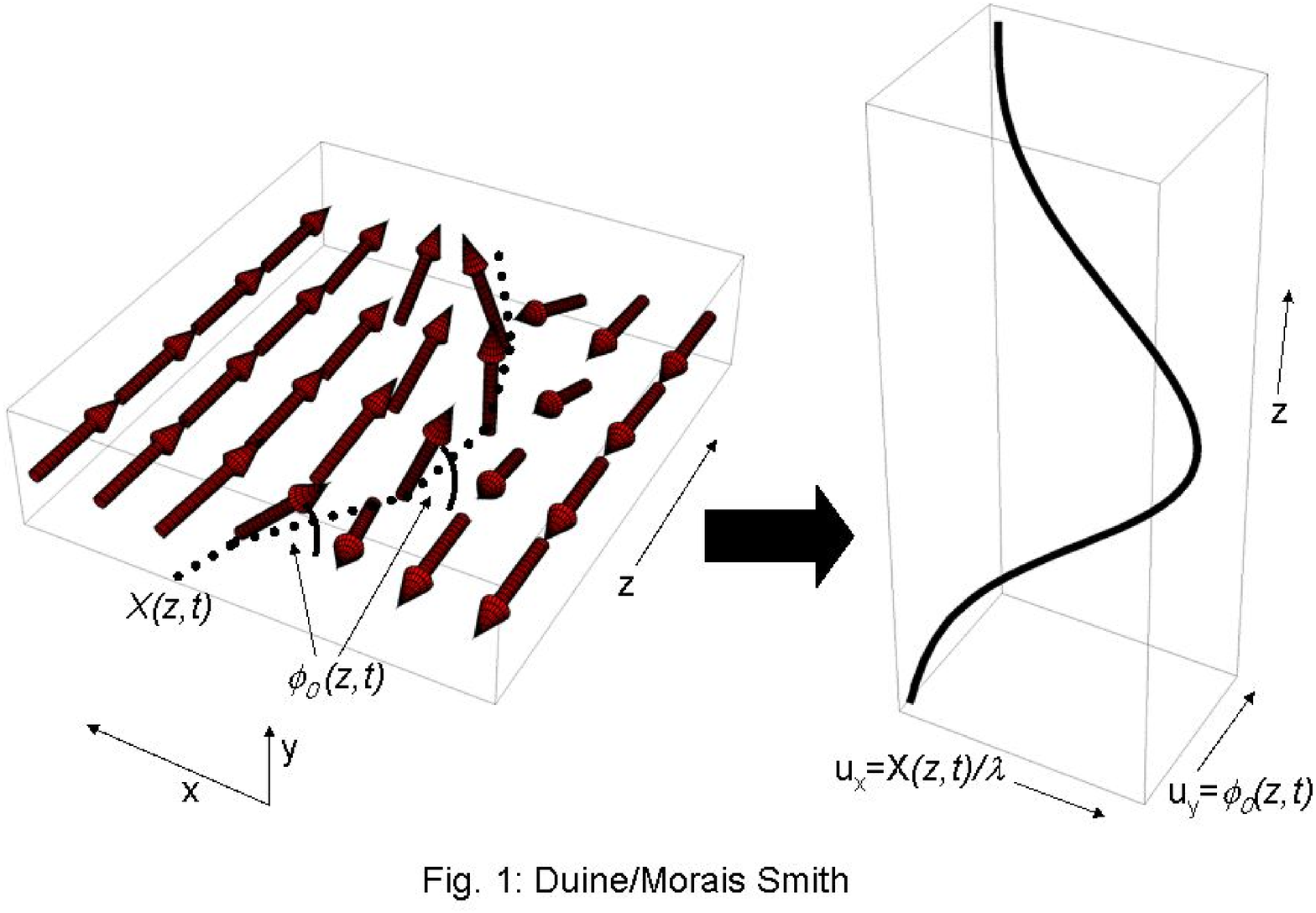,
width=10cm}}
 \caption{Mapping of current-driven domain wall dynamics to that of a vortex line. The position of the domain wall $X(z,t)$ and its
 chirality $\phi_0 (z,t)$ become the position $(u_x,u_y)$ of the vortex  via $(u_x,u_y)\equiv(X/\lambda,\phi_0)$. The potential landscape for this vortex
 is in general anisotropic. In particular, the tilting in the $u_x$-direction is set by the external magnetic field and the dissipative spin transfer torque.
 The tilting in the $u_y$-direction is determined by the reactive spin transfer torque.}
 \label{fig:mapping}
\end{figure}

In this paper we present a model for a current-driven {\it
elastic} domain-wall line moving in {\it one dimension} in the
presence of {\it disorder} and {\it thermal fluctuations}. A
crucial ingredient in the description of current-driven motion is
the chirality of the domain wall, which acts like an extra degree
of freedom. This enables a reformulation of current-driven domain
wall motion as a vortex line moving in an anisotropic potential in
{\it two dimensions} (see Fig.~\ref{fig:mapping}), which we
present in detail in Sec.~\ref{sec:dwasvortexline}. Using this
physical picture, we analyze in Sec.~\ref{sec:dwcreep} the
different regimes of pinning within the framework of collective
pinning theory. \cite{blatter1994} We present results on the
velocity of the domain-wall line as a function of current, both in
the regime where intrinsic pinning due to magnetic anisotropy
dominates, and in the extrinsic-pinning-dominated regime. Finally,
we discuss in Sec.~\ref{sec:disc} our theoretical results in
relation to recent experiments on current-driven domain walls in
GaMnAs. \cite{yamanouchi2007} Although these experiments remain,
in our opinion, not fully understood, we suggest that they may be
explained by assuming a specific form of the pinning potential for
the domain-wall line. We suggest futher experiments that could
corroborate this suggestion.

\section{Domain wall as a vortex line} \label{sec:dwasvortexline}
The equation of motion
for the magnetization direction $\bm{\Omega}$ in the presence of a
transport current $j$ is, to lowest order in temporal and spatial
derivatives, given by
\begin{eqnarray}
\label{eq:LLGwithSTTs} && \left( \frac{\partial }{\partial t} +
{\bf v}_{\rm s} \cdot \nabla  \right) \bm{\Omega} - \bm{\Omega}
\times \left( {\bf H} +{\bf H}_{\rm ext}+ \bm{\eta} \right) = \nonumber \\
 && - \alpha_G \bm{\Omega} \times  \left( \frac{\partial }{\partial
 t}+ \frac{\beta}{\alpha_G} {\bf
v}_{\rm s} \cdot \nabla \right) \bm{\Omega}~.
\end{eqnarray}
The left-hand side of this equation contains the reactive
\cite{footnote1} spin transfer torque \cite{bazaliy1998}
proportional to the velocity ${\bf v}_{\rm s}=Pj/(e\rho_{\rm s})$.
The latter velocity characterizes the efficiency of spin transfer.
Here, $P$ is the polarization of the current in the ferromagnet,
$e$ is the carrier charge, and the spin density is denoted by
$\rho_{\rm s} \equiv 2/a^3$ with $a$ the lattice constant. The
other terms on the left-hand side of Eq.~(\ref{eq:LLGwithSTTs})
describe precession around the external field ${\bf H}_{\rm ext}$
and the effective field $
  {\bf H} = -\delta E[\bm{\Omega}]/(\hbar\delta \bm{\Omega})$, which is
  given by a functional derivative of the energy functional
$E[\bm{\Omega}]$ with respect to the magnetization direction. The
stochastic magnetic field $\bm{\eta}$ incorporates thermal
fluctuations, and it has zero mean and correlations determined by
the fluctuation-dissipation theorem \cite{brown1963}
\begin{equation}
\label{eq:noisecorrsfull}
  \langle \eta_\sigma (\bx,t) \eta_{\sigma'} (\bx',t') \rangle
  = \frac{2 \alpha_G k_{\rm B}T}{\hbar} \delta (t-t') a^3 \delta
  (\bx-\bx') \delta_{\sigma\sigma'}~.
\end{equation}
It can be shown that the above still holds in the presence of
current, at least to first order in the applied electric field
\cite{duine2007} that drives the transport current. The
fluctuation-dissipation theorem also ensures that in equilibrium
the probability distribution for the magnetization direction is
given by the Boltzmann distribution ${\mathcal P}[\bm{\Omega}]
\propto \exp \{ - E[\bm{\Omega}]/k_{\rm B} T\}$. The right-hand
side of Eq.~(\ref{eq:LLGwithSTTs}) contains only dissipative
terms. The Gilbert damping term is proportional to the damping
parameter $\alpha_G$, and the dissipative \cite{footnote1} spin
transfer torque is characterized by the dimensionless parameter
$\beta$. \cite{zhang2004}

We consider a ferromagnet with magnetization direction
$\bm{\Omega}  = (\sin \theta \cos \phi , \sin \theta \sin \phi,
\cos \theta)$ that depends only on the the $x$ and $z$-direction.
In addition, we take the current in the $x$-direction and the
external magnetic field in the $z$-direction. The size of the
ferromagnetic film in the $\alpha$ direction is denoted by
$L_\alpha$ ($\alpha \in \{ x,y,z\}$) and we assume that $L_y \ll
L_z$. The latter assumption allows us to model the domain wall as
a line. Furthermore, we take the ferromagnet to have an easy
$z$-axis and hard $y$-axis, with anisotropy constants $K$ and
$K_\perp$, respectively. The spin stiffness is denoted by $J$.
With these assumptions, static domain walls have a width
$\lambda=\sqrt{J/K}$, and are, for the simplest model to be
discussed in more detail below [see
Eq.~(\ref{eq:energyfctclean})], described by the solutions
$\theta_{0} (\bx) = \cos^{-1} \left[ \tanh \left( x/\lambda\right)
  \right]$ and $\phi (\bx) =0$. To arrive at a description of
the dynamics of the domain wall, we use two collective coordinates
which may depend on the $z$-coordinate, so that the domain wall is
modelled as a line. The collective coordinates are the position of
the wall $X (z,t)$ and the chirality $\phi_0 (z,t)$. The latter
determines the sense in which the magnetization rotates upon going
through the domain wall.  The result of
Ref.~[\onlinecite{duine2006}] is straightforwardly generalized to
the case of a domain wall line. This amounts to solving
Eq.~(\ref{eq:LLGwithSTTs}) variationally with the {\it ansatz}
$\theta_{\rm dw} (\bx,t) = \theta_0 \left((x-
X(z,t))/\lambda\right)$ and $ \phi_{\rm dw} (\bx,t) = \phi_0
(z,t)$, which yields the equations of motion
\begin{eqnarray}
\label{eq:eomdwgeneral}
  \frac{\partial \phi_0}{\partial t} + \frac{\alpha_G}{\lambda} \frac{\partial X}{\partial t}
   &=& \frac{-a^3}{2\hbar L_y\lambda} \frac{\delta V }{\delta X} + \frac{\beta v_{\rm s}}{\lambda} - H_{\rm ext} +
   \eta_X (z,t)~; \nonumber \\
   \frac{1}{\lambda} \frac{\partial X}{\partial t} - \alpha_G
   \frac{\partial \phi_0}{\partial t} &=&
   \frac{a^3}{2\hbar L_y\lambda^2}\frac{\delta V }{\delta \phi_0} + \frac{v_{\rm
   s}}{\lambda}+
   \eta_{\phi} (z,t)~,
\end{eqnarray}
where the domain-wall energy
\begin{equation}
V[X,\phi_0]\equiv E[\theta_{\rm dw},\phi_{\rm dw}]~,
\end{equation}
and the stochastic forces are determined from
\begin{eqnarray}
\label{eq:noisecorrsvariational}
  && \langle \eta_\phi (z,t) \eta_\phi (z',t') \rangle
  = \langle \eta_X (z,t) \eta_X (z',t') \rangle
  \nonumber \\&&= \left( \frac{\alpha_G  k_{\rm B} T}{\hbar}\right) \left(\frac{a^3}{\lambda^2
  L_y}\right)
    \delta \left( \frac{z-z'}{\lambda}\right) \delta (t-t')~.
\end{eqnarray}
The above equations are derived using a variational method for
stochastic differential equations based on their path-integral
formulation. \cite{duine2006,duine2002} Their validity is
confirmed {\it a posteriori} by noting that in equilibrium the
probability distribution function for the position and chirality
of the domain wall is the Boltzmann distribution. That is, the
Fokker-Planck equation for the probability distribution ${\mathcal
P}[X,\phi_0]$ of the domain wall position and chirality that
follows from
Eqs.~(\ref{eq:eomdwgeneral})~and~(\ref{eq:noisecorrsvariational}),
is given by \cite{riskenbook}
\begin{eqnarray}
 && \left( 1+\alpha^2_G\right) \frac{2\hbar L_y\lambda^2}{a^3} \frac{\partial {\mathcal P}[X,\phi_0]}{\partial
 t} = \nonumber \\
 && \int dz \frac{\delta}{\delta X(z)}\left(  \alpha_G  \lambda \frac{\delta V }{\delta X(z)}
 -   \frac{\delta V }{\delta \phi_0 (z)}
 \right){\mathcal P} [X,\phi_0]\nonumber \\
 &&+ \int  \frac{dz}{\lambda}\frac{\delta}{\delta \phi_0 (z)}\left(  \lambda  \frac{\delta V }{\delta X (z)}
 +  \alpha_G \frac{\delta V }{\delta \phi_0 (z)}
 \right){\mathcal P} [X,\phi_0] \nonumber \\
 && +\alpha_G  k_B T \int \frac{dz}{\lambda}
 \left( \frac{\delta^2}{\delta \phi_0^2 (z)}+\lambda^2 \frac{\delta^2}{\delta X^2 (z) }\right)
 {\mathcal P }[X,\phi_0]~.
\end{eqnarray}
Upon insertion of the Boltzmann distribution ${\mathcal P}_{\rm
eq} [X,\phi_0] \propto \exp \left\{ - V[X,\phi_0]/(k_{\rm B} T)
\right\}$ into this equation, one straightforwardly verifies that
it is indeed a stationary solution.

Rewriting the equations of motion for the domain wall position and
chirality in terms of the dimensionless coordinate ${\bf u} (z,t)
\equiv (X(z,t)/\lambda,\phi_0 (z,t))$, we find from
Eq.~(\ref{eq:eomdwgeneral}) that the domain wall is described by
\begin{equation}
\label{eq:vortexeom}
  \epsilon_{\alpha\alpha'} \dot u_{\alpha'}  (z,t)= - \alpha_G \dot u_\alpha (z,t) -
  \frac{\delta \tilde V[{\bf u}]}{\delta u_\alpha (z,t)} + \eta_\alpha
  (z,t)~,
\end{equation}
with $\epsilon_{\alpha\alpha'}$ the two-dimensional Levi-Civita
symbol. (Summation over repeated indices $\alpha,\alpha' \in
{x,y}$ is implied. Note that $\eta_\alpha=\eta_{X,\phi}$ for
$\alpha=x,y$. ) The above equation of motion
[Eq.~(\ref{eq:vortexeom})] corresponds to the overdamped limit of
vortex-line dynamics in an anistropic potential $\tilde V[{\bf
u}]$. The left-hand side of Eq.~(\ref{eq:vortexeom}) corresponds
to the Magnus force on the vortex. We emphasize that a mass term
is missing, indicating we are indeed dealing with the overdamped
limit of vortex motion. (Note that the mass of the fictitious
vortex is not related to the D\"oring domain wall mass
\cite{doering} that arises from eliminating the chirality from the
domain-wall description, which is valid provided the latter is
small. \cite{braun1996} As the dynamics of the domain-wall
chirality is essential for current-driven domain wall motion, this
latter approximation is not sufficient for our purposes.) The
right-hand side of the equation of motion contains a damping term
proportional to $\alpha_G$, and a term representing thermal
fluctuations. The force is determined by the potential
\begin{equation}
\label{eq:vortexpot}\tilde V= \frac{a^3 V[\lambda u_x,u_y]}{2\hbar
L_y\lambda^2}+ \int \frac{dz}{\lambda} \left[\left(\frac{\beta
v_{\rm s}}{\lambda}-H_{\rm ext}\right) u_x + \frac{v_{\rm
s}}{\lambda} u_y \right]~.
\end{equation}
The tilting of this potential in the $u_x$-direction is determined
by the parameter $\beta$, the current $v_{\rm s}$, and the
external field $H_{\rm ext}$. The tilting in the $u_y$-direction
is determined only by the current. The model in
Eqs.~(\ref{eq:vortexeom})~and~(\ref{eq:vortexpot}), illustrated in
Fig.~\ref{fig:mapping}, is the central result of this paper. In
the following section, we obtain results from this model for the
domain wall velocity in different regimes of pinning, specializing
to the case of current-driven domain wall motion ($H_{\rm
ext}=0$).

\section{Domain wall creep} \label{sec:dwcreep}
In this section, we obtain results for the average drift velocity
of the domain wall as a function of applied current. First, we
discuss the situation without disorder, hereafter we incorporate
the effects of disorder.

\subsection{Intrinsic pinning} \label{subsec:intrpinning} In this
subsection, we make two assumptions that are not related. First,
we consider a homogeneous system, i.e., a system without disorder
potential $V_{\rm pin}=0$. Second, we take $\beta=0$. In this
case, the domain wall is intrinsically pinned. \cite{tatara2004}
This comes about as follows. For the magnetic nanowire model
discussed in the previous section, the energy functional in the
clean limit is given by
\begin{eqnarray}
\label{eq:energyfctclean}
  && E [\bm{\Omega}] = \int \frac{d\bx}{a^3}
  \left\{ \frac{J}{2}\left[\left( \nabla \theta \right)^2 + \sin^2 \theta \left(\nabla \phi\right)^2 \right]
  \right. \nonumber \\ &&
  \left. +\frac{K}{2} \sin^2 \theta + \frac{K_\perp}{2} \sin^2 \theta \sin^2 \phi
  \right\}~.
\end{eqnarray}
Upon insertion of the domain wall {\it ansatz} into the above
energy functional, we find that
\begin{equation}
\label{eq:intrinsicpot}
  \tilde V [{\bf u}] =
  \int \frac{dz}{\lambda}  \left[
    \frac{J}{2\hbar} \left(\frac{\partial {\bf u}}{\partial z}\right)^2
    - \frac{K_\perp}{4\hbar} \cos (2 u_y) + \frac{v_{\rm s}}{\lambda} u_y
  \right]~.
\end{equation}
Because the above potential is independent of $z$,  the domain
wall remains straight at zero temperature, i.e., $\partial {\bf
u}/\partial z =0$. By solving the equations of motion in
Eq.~(\ref{eq:vortexeom}) for the potential in
Eq.~(\ref{eq:intrinsicpot}) at zero temperature and for a straight
domain wall, one finds that $\langle| \dot {\bf u}| \rangle
\propto \sqrt{v_{\rm s}^2-\left(\lambda K_\perp/2\hbar \right)^2}$
so that the domain wall is pinned up to a critical current given
by $j_c = \lambda K_\perp e \rho_{\rm s}/2\hbar P$. (The brackets
$\langle \cdots \rangle$ denote time and thermal average.) This
intrinsic pinning is entirely due to the anisotropy energy,
\cite{tatara2004} determined by $K_\perp$, and does not occur for
field-driven domain-wall motion, or current-driven domain-wall
motion with $\beta \neq 0$. Physically, it comes about because,
for the model of a domain wall that we consider here, the reactive
spin transfer torque causes the magnetization to rotate in the
easy plane. This corresponds to an effective field that points
along the hard axis. Because the Gilbert damping causes the
magnetization to precess towards the effective field, the current
tilts the magnetization out of the easy plane. This leads to a
cost in anisotropy energy which stops the drift motion of the
domain wall if the current is too small. By solving the equations
of motion for the potential in Eq.~(\ref{eq:intrinsicpot}) at
nonzero temperature in the limit of a straight wall, one recovers
the result of Ref.~[\onlinecite{duine2006}].

At nonzero temperature, the domain wall is no longer straight.
Since only the chirality is important, our model for
current-driven domain-wall motion in Eq.~(\ref{eq:intrinsicpot})
then corresponds to the problem of a string in a tilted-washboard
potential, that has been studied before \cite{buetikker1981} in
different contexts. At nonzero temperature the string propagates
through the tilted-washboard potential by nucleating a
kink-antikink pair in the $z$-direction of the domain-wall
chirality $\phi_0(z,t)$. The kink and antikink are subsequently
driven apart by the tilting of the potential, which results in the
propagation of the string.

In the limit when the current is close to the critical one, a
typical energy barrier is determined by the competition between
the elasticity of the string and the tilted potential.
\cite{blatter1994} For $(j_c-j)/j_c \ll 1$ the cosine in the
energy functional in Eq.~(\ref{eq:intrinsicpot}) may be expanded
around one of its minima, which yields
\begin{eqnarray}
\label{eq:intrinsicpotlinearized}
  \tilde V [{\bf u}] &=&
  \int \frac{dz}{\lambda}  \left[
    \frac{J}{2\hbar} \left(\frac{\partial \delta u_y}{\partial z}\right)^2
   \right.  \nonumber \\
   && \left.  +\frac{K_\perp}{\hbar}\sqrt{1-\left(\frac{j}{j_c}\right)^2}
   \delta u_y^2 + \frac{2v_s}{3\lambda} \delta u_y^3
  \right]~,
\end{eqnarray}
where we have omitted an irrelevant constant. In the above
expression, $\delta u_y$ denotes the displacement from the
minimum. Note that we have dropped the dependence of the potential
on $u_x$ which is allowed because the potential is not tilted in
the $u_x$-direction (provided that $\beta=0$).

The potential in Eq.~(\ref{eq:intrinsicpotlinearized}) has a
minimum for $\delta u_y^{\rm min}=0$ (by construction) and a
maximum for $\delta u^{\rm max}_y = - v_s K_\perp
\sqrt{1-(j/j_c)^2}/\lambda \hbar$. The pinning potential energy
barrier, i.e., the pinning potential evaluated at the maximum,
scales as $\Delta V \propto [1-(j/j_c)^2]^{3/2}$. Consider now the
situation that a segment of length $L$ of the string is displaced
from the minimum and pinned by the maximum of the potential. The
length $L$ is then determined by the competition between the
elastic energy $\sim J (\delta u_y^{\rm max}/L)^2$, that tends to
keep the domain wall straight, and the pinning potential $\Delta
V$. Equating these contributions yields for the length $L$ that
\begin{equation}
 L \propto \left[ 1- \left(\frac{j}{j_c} \right)^2\right]^{-1/4}~.
\end{equation}
The typical energy barrier that thermal fluctuations have to
overcome to propagate the domain wall is then given by evaluating
Eq.~(\ref{eq:intrinsicpotlinearized}) for a segment of this
length. This yields a typical energy barrier $\propto [ 1- (j/j_c
)^2]^{5/4}$. Putting these results together and assuming an
Arrhenius law, we find that the domain wall velocity is
\begin{equation}
\label{eq:SAN}
  \ln \langle |\dot {\bf u}| \rangle
  \propto
  - \frac{1}{k_B T}
  \frac{JL_y}{a^3}\sqrt{\frac{K_\perp}{K}}\left[ 1-
  \left(\frac{j}{j_c}\right)^2\right]^{5/4}~,
\end{equation}
for $(j_c-j)/j_c \ll 1$.

In the regime of small currents $j \ll j_c$, the typical energy
barrier depends only logarithmically on the current,
\cite{buetikker1981} so that $\langle |\dot {\bf u}| \rangle
\propto j$. This latter result for the domain-wall velocity is the
same as found from a treatment of rigid domain-wall motion at
nonzero temperature \cite{duine2006} in the limit that $j \ll
j_c$. This is understood by noting that in the limit of vanishing
current the elasticity of the domain-wall line does not enter the
expression for the typical energy barrier, \cite{buetikker1981}
the so-called ``thin-wall" limit. \cite{cristiane}

\subsection{Extrinsic pinning} \label{subsec:extrpinning}
We now add extrinsic pinning, i.e., a disorder potential $V_{\rm
pin}$ to the potential in Eq.~(\ref{eq:intrinsicpot}). Following
Ref.~[\onlinecite{tatara2004}] we assume, in first instance, that
it only couples to the position of the domain wall $u_x$ and not
to its chirality $u_y$. This assumption is made mainly to simplify
the problem. Considering now the general case that also $\beta\neq
0$, we have that
\begin{eqnarray}
\label{eq:extrinsicpot}
  && \tilde V [{\bf u}] =
  \int \frac{dz}{\lambda}  \left[
    \frac{J}{2\hbar} \left(\frac{\partial {\bf u}}{\partial z}\right)^2
    - \frac{K_\perp}{4\hbar} \cos 2 u_y  \right. \nonumber \\
    && \left. +V_{\rm pin} (u_x,z) + \beta \frac{v_{\rm s}}{\lambda} u_x+ \frac{v_{\rm s}}{\lambda} u_y
 \rule{0mm}{6mm} \right]~.
\end{eqnarray}
We estimate a typical energy barrier using the collective pinning
theory. \cite{blatter1994,feigelman1989} Therefore, we assume that
we are in the regime where the pinning energy grows sublinearly
with the length of the wall, and that there exists a typical
length scale $L$ at which domain-wall motion occurs. (Note that we
take $L$ dimensionless since the coordinate ${\bf u}$ is
dimensionless.) \cite{blatter1994} The energy of a segment of this
length that is displaced is given by
\begin{eqnarray}
\label{eq:energylengthL}
  E(L) = \epsilon_{{\rm el}} \frac{u_x^2}{L}
  + \beta \frac{v_{\rm s}}{\lambda} L u_x +   \frac{v_{\rm s}}{\lambda} L u_y~.
\end{eqnarray}
The first term is the elastic energy with $\epsilon_{\rm el} =
J/2\hbar\lambda^2$. The second and third term correspond to the
dissipative and reactive spin transfer torques, respectively. Note
that since the dissipative spin transfer torque acts like an
external magnetic field we are able to incorporate it in the above
energy. The potential $V_{\rm pin} (u_x,z)$ leads to roughening in
the $u_x$ direction. Following standard practice,
\cite{blatter1994,feigelman1989,lemerle1998} we assume a scaling
law $u_x (L) = u_{x0} L^\zeta$, with $\zeta$ the equilibrium
wandering exponent, already mentioned in the introduction, and
$u_{x0}$ a constant. The displacement in the $u_y$ direction is
not roughened, because we have assumed that $V_{\rm pin} (u_x,z)$
does not depend on $u_y$, i.e., the domain-wall chirality. Rather,
the displacement in this direction is determined by the minima of
the potential in Eq.~(\ref{eq:intrinsicpot}) and we have that
$u_y= u_{y0}$ independent of $L$ for $j \ll j_c$. Note that in
this limit the elastic energy due to displacement in the
$u_y$-direction can also be neglected. \cite{blatter1994} Hence,
we find that
\begin{eqnarray}
\label{eq:energylengthLwithzeta}
  E(L) = \epsilon_{{\rm el}} u_{x0}^2 L^{2\zeta-1}
  + \beta \frac{v_{\rm s}}{\lambda} u_{x0} L^{\zeta+1} +   \frac{v_{\rm s}}{\lambda} L u_{y0}~.
\end{eqnarray}
Minimizing this expression with respect to $L$ then leads to a
typical energy barrier. Assuming an Arrhenius law,
\cite{blatter1994,feigelman1989,lemerle1998} we find for the
domain wall velocity
\begin{equation}
\label{eq:extrinsiccreep}
  \ln \langle |\dot {\bf u}| \rangle \propto
  - \frac{\epsilon_{\rm el}}{k_B T} \left( \frac{j_c}{j}
  \right)^{\mu_c}~.
\end{equation}
For $\beta=0$ we have that $\mu_c=(2\zeta-1)/(2-2\zeta)$. For
$\beta \neq 0$ we find $\mu_c=(2\zeta-1)/(2-\zeta)$. In
particular, for $\zeta=2/3$, applicable to domain walls in
ferromagnetic metals, \cite{lemerle1998} we have $\mu_c=1/2$ for
$\beta=0$, and $\mu_c=1/4$ for $\beta\neq 0$. Since the
dissipative spin transfer torque, proportional to $\beta$, acts
like an external magnetic field on the domain wall [see
Eqs.~(\ref{eq:vortexeom})], we recover the usual results for
field-driven domain wall motion \cite{lemerle1998} from our model.
This result is also understood from the fact that an external
magnetic field does not tilt the domain wall potential in the
chirality direction, as opposed to a current, so that the domain
wall chirality plays no role in field-driven domain wall creep. We
note that Eq.~(\ref{eq:eomdwgeneral}), or, equivalently,
Eq.~(\ref{eq:vortexeom}), contains a description of Walker
breakdown \cite{schryer1974} in the clean zero-temperature limit
and is also able to describe the transition from the creep regime
to the regime of precessional field-driven domain-wall motion
observed recently. \cite{metaxas2007}

\section{Discussion and conclusions} \label{sec:disc}
In very recent experiments on domain walls in the ferromagnetic
semiconductor GaMnAs, Yamanouchi {\it et al.},
\cite{yamanouchi2007} have observed field-driven domain-wall creep
with exponent $\mu_f \simeq 1$, and current-driven creep with
$\mu_c \simeq 1/3$, over $5$ orders of magnitude of domain-wall
velocities. The fact that these two exponents are different could
imply that $\beta$ is extremely small for this material. For
$\beta=0$ and the specific pinning potential discussed in the
previous section it is, however, impossible to find a single
roughness exponent that yields both $\mu_f=1$ and $\mu_c=1/3$.
(Note that the theoretical arguments in
Ref.~[\onlinecite{yamanouchi2007}] give $\mu_f=1$ and
$\mu_c=1/2$.)

Although it is extremely hard to determine the microscopic
features of the pinning potential, we emphasize that if pinning is
not provided mainly by {\it point-like} defects (as considered in
this paper and argued by Yamanouchi {\it et al.} to be the case in
their experiments, \cite{yamanouchi2007}) but consists of random
{\it extended} defects, the creep exponents would change
dramatically. Indeed, the latter type of disorder, which could
occur in samples if there are, e.g., steps in the height of the
film, allows for a variable-range hopping regime for creep in
which the exponent $\mu =1/3$ in the two-dimensional case.
Moreover, upon increasing the driving force, a crossover occurs to
the so-called half-loop regime where the exponent $\mu=1$.
\cite{blatter1994,nelson1993} An alternative explanation for the
experimental results of Yamanouchi {\it et al.}
\cite{yamanouchi2007} would be that $\beta \neq 0$ so that the
behavior for field and current-driven motion is similar. If the
pinning potential is random and extended, it would be possible
that the current-driven experiment is probing the variable-range
hopping regime with $\mu=1/3$, whereas the field-driven case
probes the half-loop regime with $\mu=1$. This scenario would also
reconcile the results of Ref.~[\onlinecite{yamanouchi2007}] with
previous ones \cite{yamanouchi2006} which yielded a critical
exponent of $\mu \simeq 0.5$, as the latter could be in a
different regime of pinning. In conclusion, further experiments
are required to clarify this issue. The conjecture of pinning by
extended defects may be experimentally verified by increasing the
driving in the current-driven case and checking if the exponent
crosses over from $\mu=1/3$ to $\mu=1$, while remaining in the
creep regime. Finally, since the exponent $\mu=1/3$ occurs
strictly for variable-range hopping in two dimensions, we note
that the mapping presented in this paper is crucial in obtaining
this result.

\acknowledgements

It is a great pleasure to thank G. Blatter, J. Ieda, S. Maekawa,
and H. Ohno for useful remarks.

\end{document}